\documentclass[a4paper]{article}
\usepackage{preamble}
\usepackage[
	citestyle=numeric,
	style=numeric,
	backend=biber, 
	sorting=none]{biblatex}
\title{Computing the interaction of light pulses with objects moving at relativistic speeds}
\author[1]{Maxim Vavilin\footnote{maxim@vavilin.de}}
\author[1]{Juan Diego Mazo--V\'asquez}
\author[2]{Ivan Fernandez-Corbaton}
\affil[1]{Institute of Theoretical Solid State Physics, Karlsruhe Institute of Technology, 76128, Karlsruhe, Germany}
\affil[2]{Institute of Nanotechnology, Karlsruhe Institute of Technology, 76021, Karlsruhe, Germany}
\date{}

\begin{document}
\maketitle

\begin{abstract}
	The interaction of light with short light pulses is relevant in optical traps, optical tweezers, and many other applications. The theoretical description of such polychromatic light-matter interaction is challenging, and more so when the object is moving with respect to the light source, albeit with constant speed. Light sails are futuristic examples where such speed should reach the relativistic regime. In here, we provide a methodology for the theoretical and numerical analysis of the interaction of light pulses with objects moving with constant speed. The methodology allows one, in particular, to readily compute the transfer of fundamental quantities such as energy and momentum from the light pulse to the object. As an example, we compute the transfer of energy and momentum between a given pulse and a silicon sphere moving at relativistic speeds. The methodology, however, is valid for generic pulses and objects. Particularizing the equations to the case of zero speed allows one to treat static or quasi-static objects. The method is based on the polychromatic T-matrix formalism, which leverages the many publicly available resources for computing T-matrices. 
\end{abstract}

\newpage

\section{Introduction and summary}
The manipulation of matter using light has been an active area of interest in optics and photonics for decades \cite{Ashkin2000}. The transfer of momentum and angular momentum from light to matter allows one to exert control over objects, influencing their motion in optical traps and optical tweezers, which sometimes use short light pulses \cite{Ambardekar2005}. Light pulses are also used to measure particle sizes and refractive indexes \cite{bech2010}, and even to change intrinsic material properties such as magnetization \cite{stanciu2007}. Some futuristic proposals aim at using lasers to propel light sails deep into space \cite{Daukantas2017, parkin2018, kudyshev2021}. Many of these applications pose a theoretical challenge because the calculation of the mechanical effects of light-matter interaction is very commonly done in the monochromatic regime, which covers neither light pulses nor the frequency changes experienced by light when it interacts with moving objects.

In this article, we provide a methodology for the theoretical and numerical analysis of the interaction of light pulses with moving objects. For this, we use the recently developed polychromatic T-matrix formalism \cite{Vavilin2023}. It extends Waterman's monochromatic T-matrix formalism \cite{waterman1965} to the polychromatic domain, allowing one to describe scenarios where the light-matter interaction mixes the frequencies of the incident spectrum. The T-matrix of an object is an operator that maps incident fields to the corresponding scattered fields, and fully describes the linear interaction between light and the object. The polychromatic T-matrix method provides a natural way to treat the interaction of light with objects moving with constant speed because the method is based on the \Poincare group, the symmetry group of the Minkowski space-time which includes Lorentz boosts. The Lorentz boosts are transformations that change the frame of reference to a one moving with constant speed. Similarly, Lorentz boosts may be used to describe moving objects. The formalism also gives access to the scalar product that can be used to efficiently compute the transfer of quantities such as energy and momentum between the field and the object. The proposed methodology applies to generic objects, unlike earlier methods that were limited to the relativistic motion of spherical objects \cite{Handapangoda2011, Garner2017,Whittam2023}. Given that the polychromatic domain supports the linear dependence between general time-dependent incident and scattered fields, the polychromatic T-matrix is suitable for describing time-dependent scattering in general, and hence it complements existing approaches \cite{kristensson2016, martin2021}. 

The rest of the article is organized as follows. In Sec.~\ref{sec:tmat}, the elements of the polychromatic T-matrix formalism that are essential for this work are shortly explained: the wave function, the electromagnetic scalar product, and the Lorentz boost of electromagnetic fields. In Sec.~\ref{sec:transfer}, the formulas and the procedure for analyzing the interaction between a generic light pulse and a generic object moving at a constant speed are introduced. As an example, the interaction of a particular light pulse and a relativistically moving silicon sphere is investigated for a wide variety of speeds, with an emphasis on the transfer of energy and momentum between the light pulse and the sphere. The analysis is conducted in both co-moving and laboratory frames. We establish that the quantities in the laboratory frame should be obtained by transforming the ones obtained in the object frame, instead of doing the analysis directly in the moving frame. While we provide the formulas that theoretically allow one to compute the T-matrix of a moving object, we also demonstrate that, in practice, the multipolar orders of the light-matter interaction cannot really be truncated without losing information about the scatterer. Section~\ref{sec:concl} concludes the article.

The presented theoretical framework can help in advancing the technology of devices such as light sails, including the practical optimization of sail designs and material selection to enhance the efficiency of the propulsion of the devices. Additionally, it can be used for predicting spectroscopic signals of objects in relativistic motion, such as, for example, chiral molecules in outer space \cite{Whittam2023b}. Moreover, the application of the methodology for objects at rest provides a rigorous and simple, yet powerful approach to the study and engineering of light-matter interaction between generic light pulses and generic objects, with clear applications to pulsed optical traps and tweezers.

The methodology is computationally friendly. The T-matrix is a popular computational strategy \cite{Gouesbet2019,Mishchenko2020}, and there are multiple algorithms to compute T-matrices of physical scatterers \cite{ganesh2010, Fruhnert2017}. There is also a variety of publicly available resources for computing T-matrices \cite{TmatrixCodes}. Together with the formulas in this paper, such resources allow one to readily compute interaction of light with relativistically moving scatteres. The code that provides the numerical results of this study is available from the corresponding author, upon reasonable request. The code uses the recently released \textsc{treams} package \cite{Beutel2023b}.

%The method can also be applied as an approximation to object moving with non-constant speed, if the rate of change of speed is small compared to the pulses' lenght.

\section{Theoretical Framework}\label{sec:tmat}
Here, we shortly summarize the relevant parts of the recently developed polychromatic T-matrix formalism \cite{Vavilin2023}, as well as the notion of the electromagnetic scalar product and the transformation rules for the electromagnetic field upon Lorentz boosts.
\subsection{Wave function and scalar product}
The total electromagnetic field outside of a sphere enclosing a scatterer may be decomposed into the regular incident and the irregular scattered fields \cite{mishchenko2002}. The polychromatic incident electromagnetic field $\bm E^\text{inc}(\bm r, t)$ and can be represented using its decomposition into multipolar fields
\eq{
	\bm E^\text{inc}(\bm r, t) &= \int_0^{\infty} dk \, k \sum_{\lambda=\pm 1} \sum_{j=1}^{\infty}  \sum_{m=-j}^j \, f_{j m \lambda}(k) \bm R_{jm\lambda}(k, \bm r, t), \label{eq:measure}
}
with speed of light $c$, wavenumber $k = \omega / c$, helicity (circular polarization) $\lambda=\pm 1$, total angular momentum $j = 1$ (dipolar fields), $j=2$ (quadrupolar fields) etc., and angular momentum along the $z$-axis $m=-j, -j+1, \dots j$. We use the complex-valued electric field $\bm E(\bm r, t)$, connected to the real field via ${\bm{\mathcal E}}( \bm r, t) = 2 \Re[\bm E(\bm r, t)]$.

The complex coefficient function of the decomposition $f_{j m \lambda}(k)$ is the wave function of the field in the angular momentum basis and the basis vector fields are
\eq{
	\bm R_{j m\lambda}(k, \bm r, t) 	&= \sqrt{\frac{c\hbar}{\epsilon_0}} \frac{k \, e^{-i kc t}}{\sqrt{\pi}\sqrt{2j+1}}  \sum_{L=j-1}^{j+1} \sqrt{2L + 1}\, i^{L} j_{L}(k r) \, C^{j\lambda}_{L0,1\lambda} \bm Y^L_{j m}(\hat{\bm r}) 
	 \label{eq:defR},
}
with reduced Plank's constant $\hbar$, permittivity of vacuum $\epsilon_0$, spherical Bessel functions $j_L(x)$, Clebsch-Gordan coefficients $C_{j_1m_1,j_2m_2}^{j_3m_3}$, and vector spherical harmonics $\bm Y^L_{j m}(\hat{\bm r})$ as defined in \cite{varshalovich1988}. The basis vector fields $\bm R_{j m\lambda}(k, \bm r, t)$ are connected to the usual electric ($\mathbf{N}$) and magnetic ($\mathbf{M}$) multipolar fields via
\eq{
\bm R_{j m\lambda}(k, \bm r, t) &= - \sqrt{\frac{c\hbar}{\epsilon_0}} \frac{1}{\sqrt{2\pi}} \, k \, i^j  \Big(  e^{-i kc t}\, \bm N_{jm}(k r, \hat{\bm r}) + \lambda \,e^{-i kc t} \, \bm M_{jm}(k r, \hat{\bm r} ) \Big) \label{eq:rmn}
}

The extra factor of $k$ in the integration measure of \Eq{eq:measure} and in the definition of Eqs.~(\ref{eq:defR},\ref{eq:rmn}) are necessary parts of the formalism that follow from the invariant scalar product and ensure the unitarity of Lorentz transformations \cite{Vavilin2023}. 

One of the advantages of using the language of wave function is the access to the invariant scalar product between two free solutions of Maxwell equations \cite{gross1964}, which, with our conventions reads:
\eq{
\braket{f|g} = \sum_{\lambda=\pm 1} \int_0^{\infty} dk\, k \sum_{j=1}^{\infty} \sum_{m=-j}^j \, f^*_{j m \lambda}(k) g_{jm\lambda}(k),\label{eq:scalarAM}
}
which has an equivalent expression in the plane wave basis:
\eq{
	\braket{f|g} = \sum_{\lambda=\pm1} \int \frac{d^3 \bm k}{k} \, f^*_\lambda (\bm k) g_\lambda (\bm k). \label{eq:scalarPW}
}
The wave function coefficients in both bases are connected via the Wigner D-matrix:
\eq{
f_\lambda(\bm k) &=  \sum_{j=1}^{\infty} \sum_{m=-j}^j \sqrt{\frac{2j+1}{4\pi}} D^j_{m\lambda}(\phi,\theta,0)^* \, f_{j m \lambda}(k),
}
where $\theta$ and $\phi$ are polar and azimuthal angles of $\bm k$, respectively, namely $\theta=\arccos\left(k_z/k\right)$, and $\phi=\text{arctan2}\left(k_y,k_x\right)$.

The known action of the symmetry generators on the wave function \cite{tung1985}, gives access to physical quantities contained in the field, for example number of photons $\braket{f|f}$, energy $\braket{f|H|f}$ and momentum in the z-direction $\braket{f|P_z|f}$ \cite[\S 9, Chap.~3]{Birula1975}. Those will be used in Sec.~(\ref{sec:transfer}), and are easy to compute numerically.

The scattered electromagnetic field can also be connected to this formalism, via the decomposition into irregular multipolar fields
\eq{
	\bm E^\text{sca}(\bm r, t) &= \int_0^{\infty} dk \, k \sum_{\lambda=\pm 1} \sum_{j=1}^{\infty}  \sum_{m=-j}^j \, g_{j m \lambda}(k) \bm S_{jm\lambda}(k, \bm r, t),  
}
where $\bm S_{jm\lambda}(k, \bm r, t)$ is defined similarly to $\bm R_{j m\lambda}(k, \bm r, t)$ with spherical Bessel functions $j_L(kr)$ substituted with spherical Hankel function of the first type halved: $\frac{1}{2} h^{(+)}(kr)$. The factor $\frac{1}{2}$ is necessary for using the same scalar product formula \Eq{eq:scalarAM} to compute quantities contained in the scattered field \cite{Vavilin2023}.

\subsection{Lorentz boost of electromagnetic field}
Lorentz boosts are transformations that relativistically describe the change to a reference frame that moves with some constant velocity. Throughout the article, unless stated otherwise, we will use their active counterpart that describes the movement of the object rather than the movement of the reference frame.

A 4-vector of a point in Minkowski space-time is transformed under a Lorentz boost in the z-direction via \cite[Chap.~10]{tung1985}
\eq{
	x^\mu = \begin{pmatrix} c t \\ x^1 \\x^2 \\ x^3 \end{pmatrix}  \mapsto L_z(\xi)^\mu_{\;\nu}\, x^\nu = 
\begin{pmatrix}
\cosh(\xi) & 0 & 0 & \sinh(\xi) \\
0 & 1 & 0 & 0 \\
0 & 0 & 1 & 0 \\
\sinh(\xi) & 0 & 0 & \cosh(\xi) \\
\end{pmatrix} \label{eq:lormatrix}
\begin{pmatrix} ct \\ x^1 \\x^2 \\ x^3 \end{pmatrix},
}
where the boost parameter is called rapidity and is connected to the velocity $v$ via $\xi = \arctanh(v/c)$. Rapidity provides a natural parametrization of boosts, making many formulas more compact, compared with formulation with $\bm v$. However, the speed of the boost will be also used in this text.

A general Lorentz boost in an arbitrary direction $\hat{\bm n}$ can be expressed as a composition of a boost in the z-direction and spacial rotations:
\eq{
	L_{\hat{\bm n}}(\xi) = R(\phi, \theta, 0) L_z(\xi) R^{-1}(\phi,\theta,0). \label{eq:bdecomp}
}
Here the direction of the boost $\hat{\bm n}$ is parametrized by polar angle $\theta=\arccos\left(n_z \right)$ and azimuthal angle $\phi=\arctantwo\left(n_y,n_x\right)$, with the rotations $R$ parametrized via Euler angles in $zyz$-convention $R(\alpha,\beta,\gamma)=R_z(\alpha)R_y(\beta)R_z(\gamma)$.

For a massless 4-wave vector $k^\mu$ with $k^\mu k_\mu=0$, or $k^0 = \abs{\bm k}\eqcolon k$, the z-direction boost implies
\eq{
k^\mu  = 
\begin{pmatrix}
\abs{\bm k} \\ k_x \\ k_y \\ k_z
\end{pmatrix}
\mapsto
\begin{pmatrix}
\cosh(\xi) \abs{\bm k} + \sinh(\xi)k_z \\ k_x \\ k_y \\ \sinh(\xi)\abs{\bm k} + \cosh(\xi)k_z
\end{pmatrix}=\tilde k^\mu. \label{eq:plboost}
}
The angles and the wave number of the boosted wave vector can be written in terms of the old ones via 
\eq{
	\cos(\tilde \theta) &= \frac{\cos(\theta)\cosh(\xi) + \sinh(\xi)}{\cosh(\xi) + \cos(\theta)\sinh(\xi)} = \frac{\cos(\theta) + \tanh(\xi)}{1 + \cos(\theta)\tanh(\xi)} \label{eq:plboosttheta} \\
	\tilde k &= k ( \cosh(\xi) + \cos(\theta) \sinh(\xi))\label{eq:plboostwn} \\
	\tilde \phi &= \phi \label{eq:plboostphi}.
}

The transformation properties of electric and magnetic fields under the actions of space-time symmetries are well-known. Specifically, the active Lorentz boost with velocity $\bm v$ of real-valued electromagnetic fields  is defined as \cite[Sec.~11.10]{Jackson1998}
\eq{
\widetilde{\bm{\mathcal E}} (\bm r, t)&=\gamma {\bm{\mathcal E}}( \tilde{\bm r}, \tilde t) -  \gamma \bm v \times {\bm{\mathcal B}}( \tilde{\bm r}, \tilde t)  - \frac{\gamma^2 \bm v}{(\gamma+1)c^2} \bm v \cdot {\bm{\mathcal E}} ( \tilde{\bm r}, \tilde t) \label{eq:lbu1}\\
\widetilde{\bm{\mathcal B}} (\bm r, t)&=\gamma {\bm{\mathcal B}}( \tilde{\bm r}, \tilde t) +  \frac{1}{c^2}\gamma \bm v \times {\bm{\mathcal E}}( \tilde{\bm r}, \tilde t)  - \frac{\gamma^2 \bm v}{(\gamma+1)c^2} \bm v \cdot {\bm{\mathcal B}} ( \tilde{\bm r}, \tilde t) \label{eq:lbu2}
}
with inversely transformed space-time point $\begin{pmatrix} c\tilde t \\ \tilde{\bm r} \end{pmatrix} =  L^{-1}(\xi) \begin{pmatrix} ct \\ \bm r \end{pmatrix}$ and $\gamma = (1-v^2/c^2)^{-1/2}$. The passive version of the Lorentz boost, i.e. the boost of the reference frame instead of the field, differs from Eqs.~(\ref{eq:lbu1}-\ref{eq:lbu2}) by the substitution $\bm v \rightarrow - \bm v$ and should not be confused with the active version. 

The Lorentz boost $L_z(\xi)$ of the wave function in plane wave basis, i.e. its change when the field is transformed under Eqs.~(\ref{eq:lbu1})-(\ref{eq:lbu2}), is \cite[Eq.~(10.4-22)]{tung1985}
\eq{
L_z(\xi) f_\lambda(\bm k) = f_\lambda(\bm k'),  \label{eq:boostpwwf}
} 
with $\bm{k}'  = L^{-1}_z(\xi) \bm k$ defined by Eqs.~(\ref{eq:plboosttheta}-\ref{eq:plboostphi}), but substituting $\xi\rightarrow -\xi$ to implement the inverse transformation.

In the angular momentum basis, the action of the boost reads \cite[Eq.~(79)]{Vavilin2023}:
\eq{
L_z(\xi) f_{jm\lambda}(k) &= \frac{1}{2}\,\sqrt{2j+1}\sum_{j'=1}^{\infty} \,\sqrt{2j'+1} \int^{1}_{-1} d(\cos \theta) \, d^j_{m\lambda}(\theta) \,  d^{j'}_{m\lambda}(\theta') f_{j' m \lambda}(k'), \label{eq:boostwfangular}
}
where the $d^j_{m\lambda}(\alpha)$ are the small Wigner-matrices as defined in \cite{tung1985}, Sec. 7.3, and with $k'$ and $\theta'$ given by
\eq{
k' &= k\big(\cosh(\xi) - \cos(\theta)\sinh(\xi)\big) \\
\cos(\theta') &= \frac{\cos(\theta) - \tanh(\xi)}{1 - \cos(\theta) \tanh(\xi)}.
}
Notably, $\lambda$ and $m$ are not changed by a boost in the $z$-direction.

\begin{figure}[h]
\makebox[\textwidth][c]{%
 \begin{subfigure}[b]{0.6\textwidth}
        \centering
        \includegraphics[width=\textwidth]{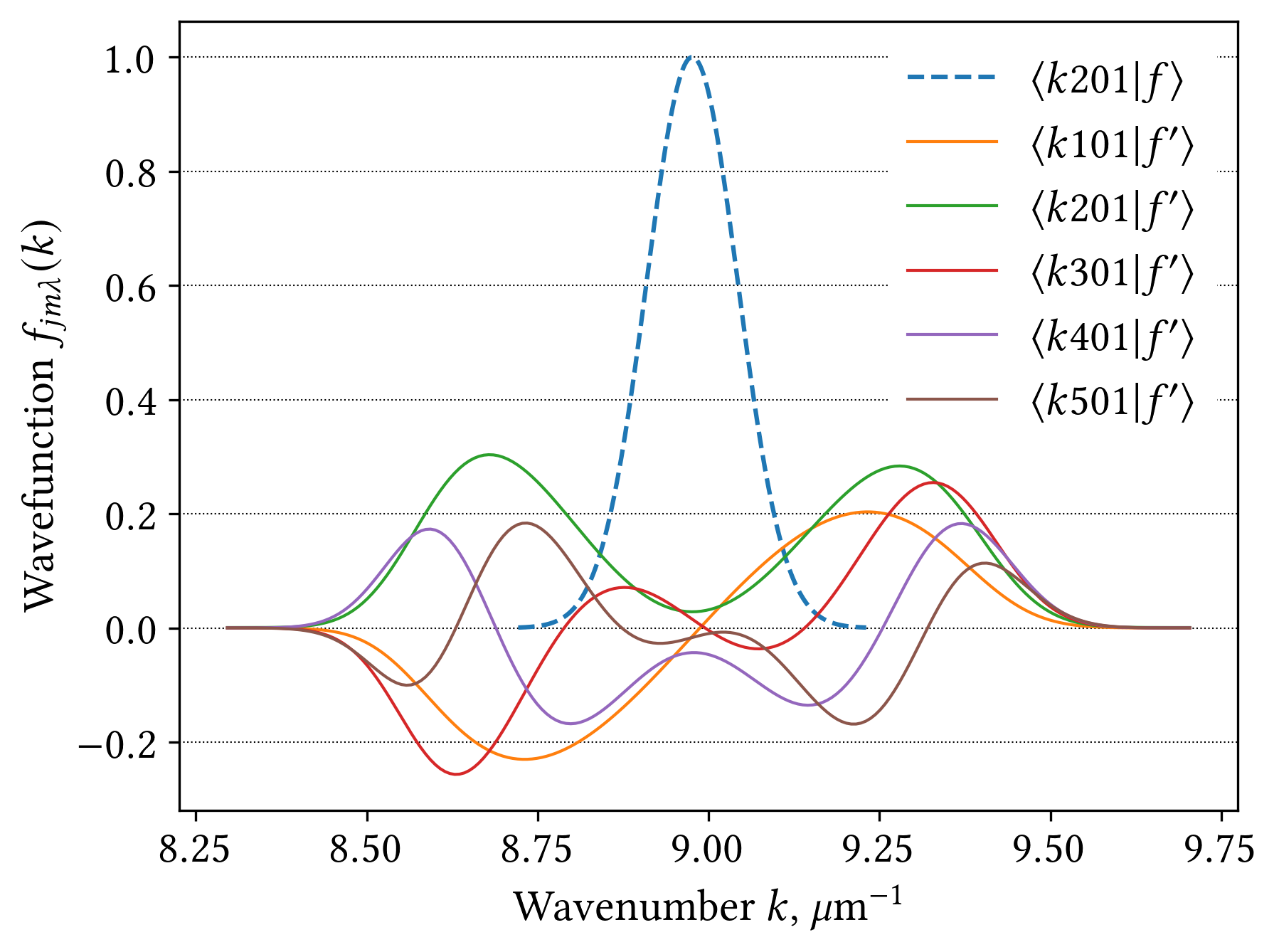}
        \caption{$\xi = 0.05$}

     \end{subfigure}
     \hfill
     \begin{subfigure}[b]{0.6\textwidth}
         \centering
        \includegraphics[width=\textwidth]{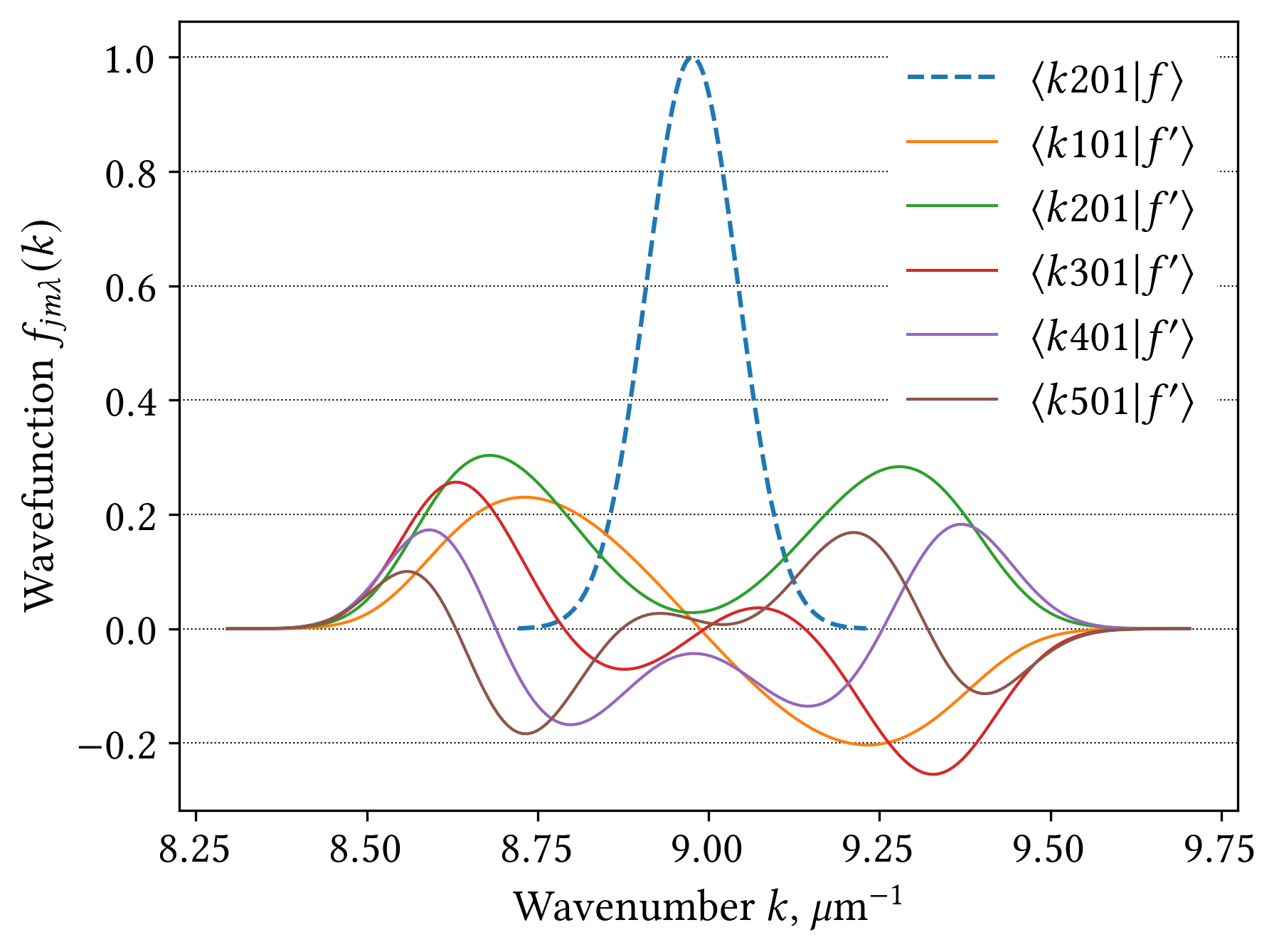}
        \caption{$\xi = -0.05$}
	
     \end{subfigure}
     }
    % \end{centering}
    % }
	\caption[Lorentz boost of an angular momentum eigenstate]{
		Active Lorentz boost of a wave function in z-direction $\ket{f'}=L_z(\xi) \ket{f}$ with positive rapidity $\xi=0.05$ (a) and negative rapidity $\xi=-0.05$ (b). The initial wave function (dashed) describes a multipolar pulse with quantum numbers $j=2$, $m=0$, $\lambda=1$, and a Gaussian spectral profile. Wave function of the boosted field $f'_{jm\lambda}(k)=\braket{kjm\lambda|f'}$ is shown for multipolar order up to $j=5$.
		}\label{fig:pwboosted}
\end{figure}

The last formula can be illustrated with an exemplary wave function of a definite angular momentum. A multipolar pulse is considered, defined by the wave function $f_{jm\lambda}(k)$ with $j=2$, $m=0$ and $\lambda=1$:
\eq{
	f_{201}(k) = e^{-\frac{(k-k_0)^2}{2 \left(\Delta_k\right)^2} }.
}
Here, $\frac{1}{c\Delta_k} = \Delta_t= 50~\si{fs}$ is a Gaussian time width such that the values of the function outside of the domain $8.72~\si{\mu\m}^{-1}\leq k \leq9.23~\si{\mu\m}^{-1}$ are insignificant. The center wavelength is $\frac{2\pi}{k_0} = 700~\si{nm}$. The coefficients $f'_{j01}(k) = \braket{kj01|f'}$ of the Lorentz boosted field are depicted in Fig.~(\ref{fig:pwboosted}). The boosted coefficients have non-zero components for all $j \in \mathbb N$, here we plot only the first five.

Apart from the change of the values of the wave function under the Lorentz boost, one may also observe the spreading of the wave function in the wavenumber domain. Note that the boosted wave function $\ket{f'}$ contains both higher and lower multipolar components than the initial $\ket{f}$.

\subsection{Polychromatic T-matrix and S-matrix}
Consider the incident and scattered parts of the total polychromatic field outside a sphere enclosing the scatterer:
\eq{
	\bm E^\text{inc}(\bm r, t) &= \int_0^{\infty} dk \, k \sum_{\lambda=\pm 1} \sum_{j=1}^{\infty}  \sum_{m=-j}^j \, f_{j m \lambda}(k) \bm R_{jm\lambda}(k, \bm r, t) \\
	\bm E^\text{sca}(\bm r, t) &= \int_0^{\infty} dk \, k \sum_{\lambda=\pm 1} \sum_{j=1}^{\infty}  \sum_{m=-j}^j \, g_{j m \lambda}(k) \bm S_{jm\lambda}(k, \bm r, t),  
}

The polychromatic T-matrix is defined as an operator that maps the wave function of the incident field onto the wave function of the scattered field:
\eq{
g_{jm\lambda}(k)& = \int_0^\infty dk' \, k' \sum_{\lambda'=\pm 1} \sum_{j'=1}^{\infty}\sum_{m'=-j'}^{j'} \, T^{jm\lambda}_{\,j'm'\lambda'}(k, k') f_{j'm'\lambda'}(k'). \label{eq:tmataction}
}
A special case of the polychromatic T-matrix is it being diagonal in frequency:
\eq{
T^{jm\lambda}_{\,j'm'\lambda'}(k, k') = \frac{1}{k} \delta(k-k')T^{jm\lambda}_{\,j'm'\lambda'}(k). \label{eq:tmatdiag}
}
This realizes a usual situation when each frequency component of the incident field contributes only to the same frequency component of the scattered field. Equation~(118) in Ref.~\cite{Vavilin2023} connects the $T^{jm\lambda}_{\,j'm'\lambda'}(k, k')$ in \Eq{eq:inout2} with the monochromatic T-matrices for the conventions used in the recently released \textsc{treams} package. 

The S-matrix is an alternative setting for formalizing light-matter interaction. The S-matrix maps incoming fields to outgoing fields. The polychromatic S-matrix\cite[Sec.~3]{Vavilin2023} is defined in our convention as, $S = \mathds{1} + T$, so the coefficients of the incoming field are equal to the coefficients of the incident field, and the outgoing part of the field is computed as
\eq{
	h_{jm\lambda}(k) &= f_{jm\lambda}(k) + (T f)_{jm\lambda}(k) \label{eq:inout2} \\
	&= f_{jm\lambda}(k) + \sum_{\lambda' = \pm 1}\sum_{j'=1}^{j_\text{max}}\sum_{m'=-j'}^{j'} T^{j m \lambda}_{j' m' \lambda'}(k) f_{j'm'\lambda'}(k),
}
where we use Eq.~(\ref{eq:tmataction}) for the case of a frequency-diagonal T-matrix \Eq{eq:tmatdiag}. 

In the S-matrix setting, the incoming and outgoing fields are the total fields before the interaction starts and after it has stopped, respectively. It is then straightforward to write down the change of a given quantity contained in the field, such as energy or momentum:
\begin{equation}
	\label{eq:transfer}
\langle \Delta\Gamma\rangle=\langle f|\Gamma|f\rangle-\langle h|\Gamma|h\rangle=\langle f|\Gamma-S^\dagger \Gamma S|f\rangle,
\end{equation}
where $\Gamma$ is the self-adjoint operator corresponding to the quantity of interest. For quantities that obey global conservation laws, such as energy, angular momentum, and linear momentum, the difference $\langle \Delta\Gamma\rangle$ is transferred to the object during the light-matter interaction. 

Equation~\ref{eq:transfer} is the basic equation that we will use to compute the transfer of energy and linear momentum from a light pulse to a relativistically moving object. For the rest of the article, we will treat the case of an achiral sphere. We highlight, however, that the methodology that we present can be applied to any object whose T-matrix is at hand. T-matrices can be obtained for virtually any material object, including individual particles of arbitrary shape, (chiral) molecules, and clusters and periodic arrangements thereof.    

\section{Interaction between a light pulse and a relativistically moving silicon sphere}\label{sec:transfer}
Here we focus on the interaction between light and a moving silicon sphere, in particular we compute the difference of energy and momentum contained in light before and after interaction with the sphere. First, the interaction is computed in the reference frame of the object (the co-moving frame), and later, in the laboratory frame. 

\subsection{Co-moving frame of reference}
\begin{figure}[h]
        \centering
	%\textbf{}
        \includegraphics[width=0.5\textwidth]{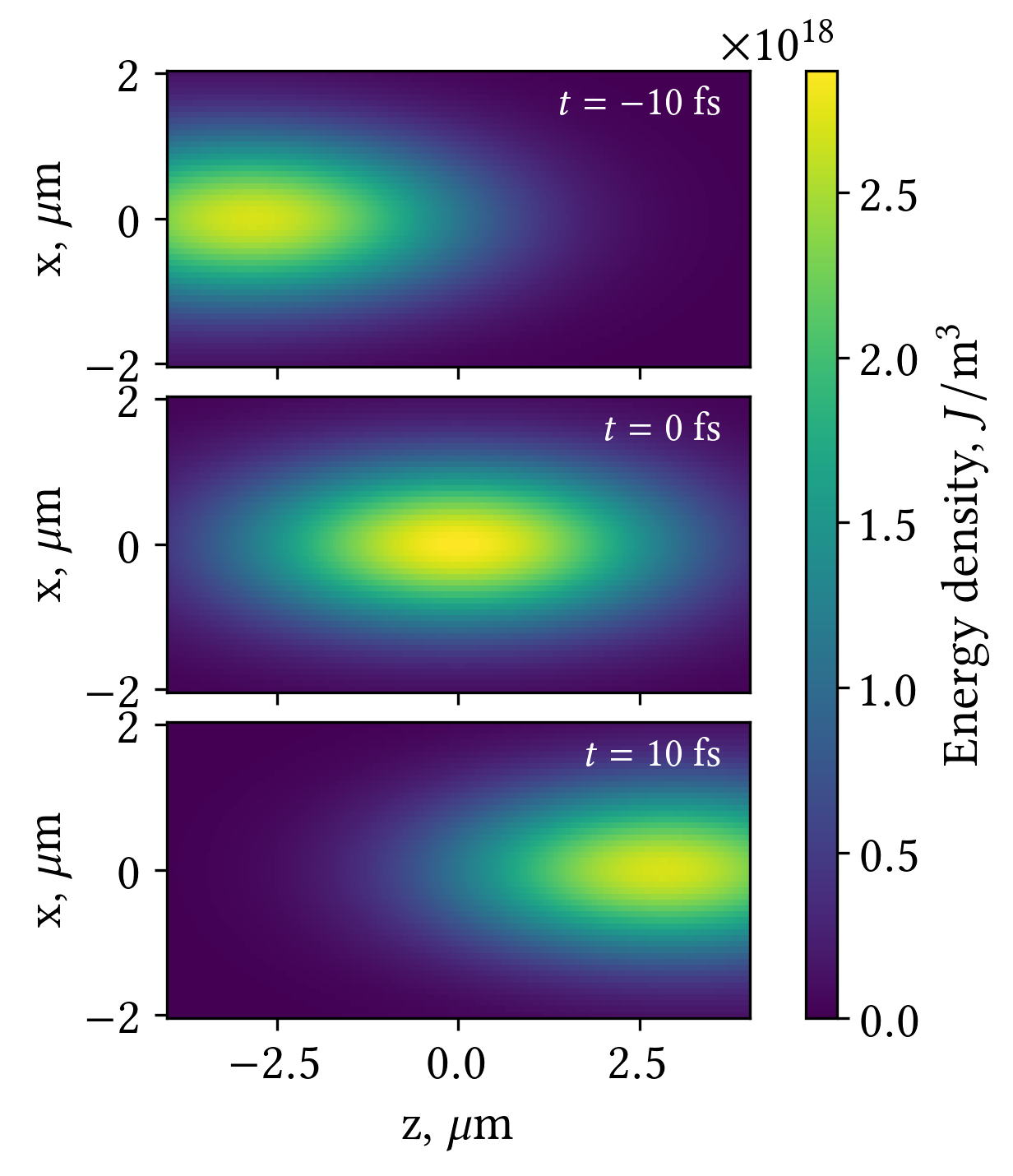}
	\caption{
		Energy density of the incident pulse in the $xz$-plane, at three different times.
		}\label{fig:Pulse}
\end{figure}

Consider an incident field $\ket f$ present in the laboratory frame, defined by the wave function in the plane wave basis as
\eq{
f_+(\bm k) &= A\, \,e^{- \frac{(k - k_0)^2 \Delta_t^2 c^2}{2}} \, e^{- \frac{\theta^2}{ 2 \Delta_\theta^2}}  e^{-i\phi }
\label{eq:apppulse}\\
f_-(\bm k) &= 0.
}
The angles $\theta$, $\phi$  are the polar and azimuthal angles of the wave vector $\bm k$. The time width of the pulse is $\Delta_t = 10~\si{\fs}$, the central wavelength $\frac{2\pi}{k_0} = 700~\si{\nm}$, and the polar angle spread $\Delta_\theta = 0.1$ radians. The constructed pulse is focused along the z-axis (see Fig.~(\ref{fig:Pulse})), propagating in the positive $z$ direction with values of $f_+(\bm k) \approx 0$ for $\theta>0.37$. The Gaussian profile makes the wave function negligible outside of the region $8.1~\si{\mu\meter}^{-1}<k<9.8~\si{\mu\meter}^{-1}$.

We set the normalization constant $A = 3.25 \times10^{11}~\si{\nm}$ to fix the energy of the pulse to 5~\si{mJ} via
\eq{
	\braket {f|H|f} = \sum_{\lambda=\pm 1} \int \frac{d^3 \bm k}{k} \abs{f_\lambda(\bm k)}^2 c \hbar k = 5 \times 10^{-3}~\si{\J},
}
where we have used that $H|\bm k \lambda\rangle=\hbar c k|\bm k \lambda\rangle$. Similarly, using that $P_z|\bm k \lambda\rangle=\hbar k_z|\bm k \lambda\rangle$, the momentum $P_z$ carried by the pulse is
\eq{
	\braket {f|P_z|f} = \sum_{\lambda=\pm 1} \int \frac{d^3 \bm k}{k} \abs{f_\lambda(\bm k)}^2 \hbar k \cos \theta = 1.66 \times 10^{-11}~\si{\kilogram\metre\per\second}.
}

We assume that the pulse interacts with a silicon sphere that moves along the z-axis at some constant speed $v=c \tanh \xi$, away or towards the pulse. In the co-moving frame the object is stationary, it is described by its frequency-diagonal T-matrix, and perceives the incident field to be Lorentz boosted in the opposite direction $|f'\rangle=L_z(-\xi) \ket f$.

We are after the quantities:
\begin{equation}
	\label{eq:transferhpz}
	\begin{split}
		\langle \Delta H\rangle^{\text{obj}}&=\langle f'|H-S^\dagger H S|f'\rangle, \text{ and }\\
		\langle P_z\rangle^{\text{obj}}&=\langle f'|P_z-S^\dagger P_z S|f'\rangle,
	\end{split}
\end{equation}
where the $^{\text{obj}}$ superscript denotes that they are computed in the reference frame co-moving with the object. In the light sail application, the energy and momentum transferred to the object in its own reference frame are crucial quantities to understand the amount of heating and acceleration caused by the pulse, respectively.

\begin{figure}[h]
 \begin{subfigure}[b]{0.5\textwidth}
        \centering
	%\textbf{}
        \includegraphics[width=\textwidth]{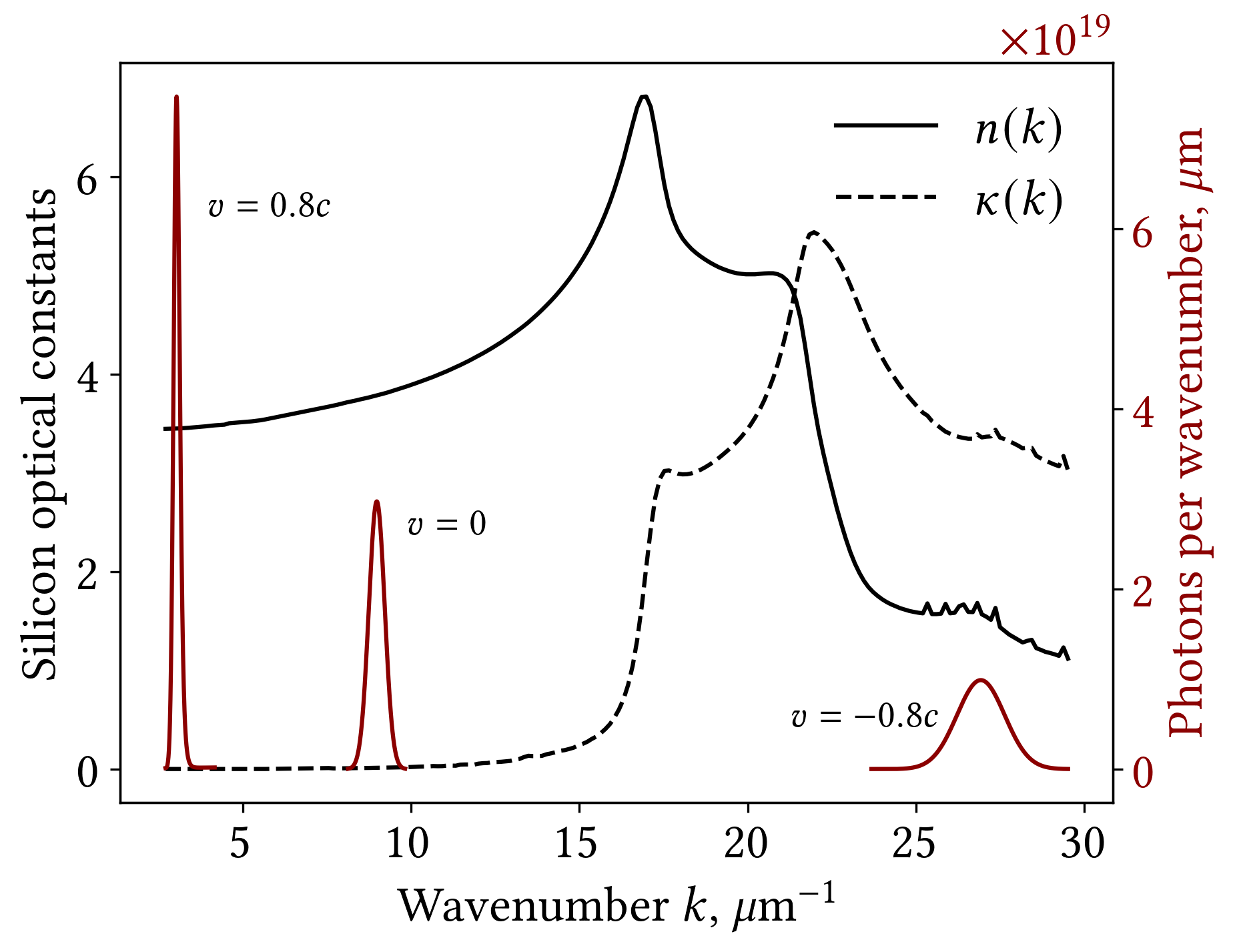}
        \caption{}
	\label{fig:optical_wide_klist}
     \end{subfigure}
     \hfill
     \begin{subfigure}[b]{0.5\textwidth}
         \centering
        \includegraphics[width=\textwidth]{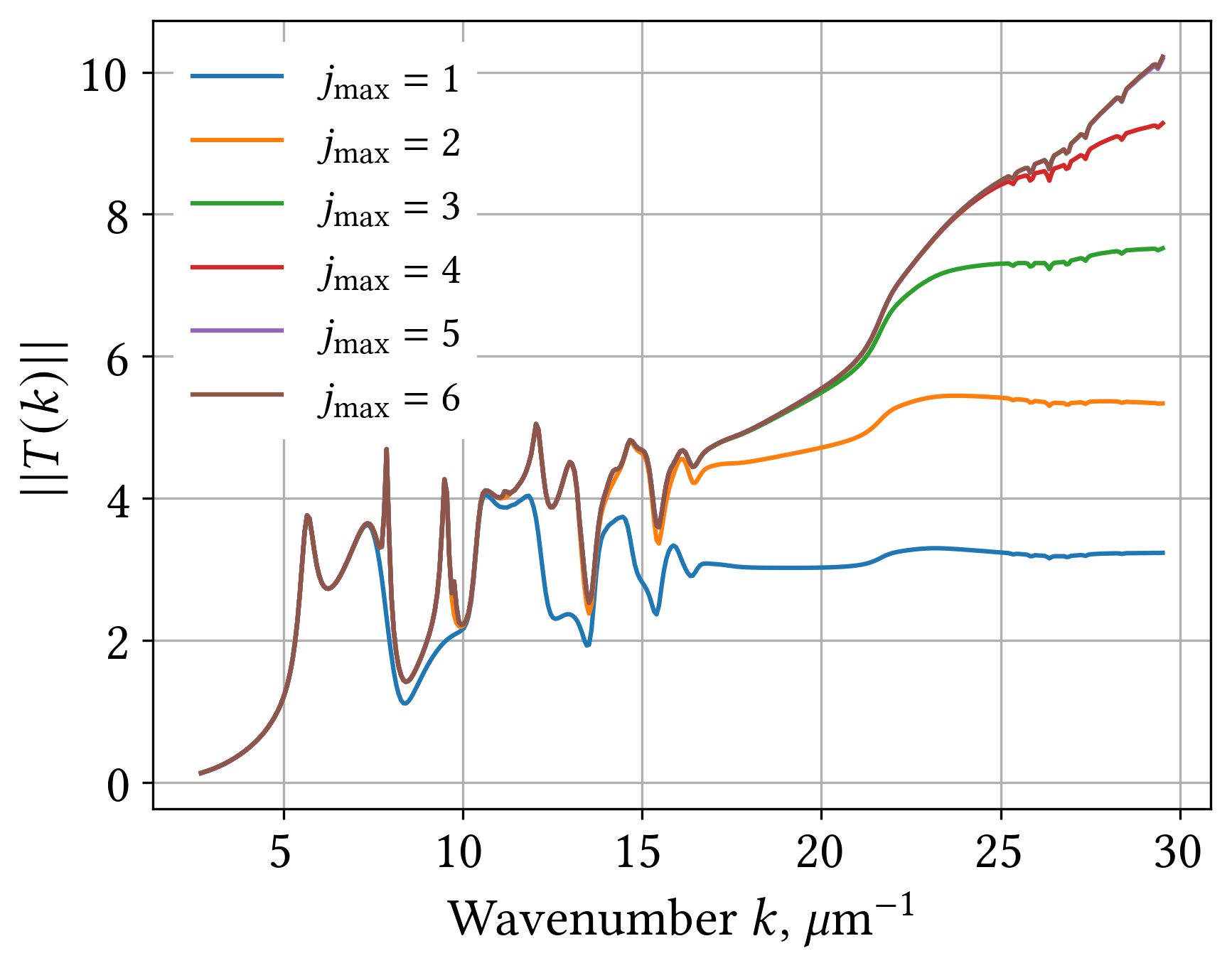}
        \caption{}
	\label{fig:crossec_wide_klist}
     \end{subfigure}
    % \end{centering}
    % }
	\caption{
		(a)  In black: refractive index $n(k)$ and extinction coefficient $\kappa(k)$ of Silicon \cite{palik1998} as functions of wavenumber on the whole domain required for the computation. In red: photon density per wavenumber of the considered pulse as perceived by the object moving with three speeds $v$. The negative sign of $v$ corresponds to the movement toward the pulse. (b) Interaction cross-section of a silicon sphere with radius 150~\si{nm} as a function of wavenumber, computed via the Frobenius norm of the monochromatic T-matrix for different maximal multipolar orders $j_\text{max}$. The brown line corresponding to $j_\text{max}=6$ completely covers the purple line of $j_\text{max}=5$, bringing no additional precision.}
\end{figure}

The numerical computations are conducted for the span of rapidities $-1.1 \leq \xi \leq 1.1$ or, equivalently, $-0.8 \leq v/c \leq 0.8$. Positive $v$ corresponds to the sphere moving in the positive z-direction, and negative $v$ to the movement in the negative direction. For each velocity the transformed field has significant components in different parts of the frequency spectrum. If the pulse in the laboratory frame is completely described on the wave number domain $k_\text{min}\leq k \leq k_\text{max}$, then the optical properties of the object should be, in general, known in the wave number region $e^{-\abs{\xi}} k_\text{min} \leq k \leq e^{\abs{\xi}} k_\text{max}$. Figure~(\ref{fig:optical_wide_klist}) depicts the optical properties of silicon \cite{palik1998} on the total wave number band required for computation in the chosen range of velocities. The maximal multipole order of the T-matrix that is required for the precise scattering simulation needs to be determined as well. Figure~(\ref{fig:crossec_wide_klist}) illustrates Frobenius norms of the monochromatic silicon T-matrices

\eq{
\norm{T(k)} =  \sqrt{\text{Trace}[T^\dagger(k) T(k)]}
} 
at different maximal multipole orders on the same wavenumber domain, motivating the truncation order to be $j_\text{max} = 5$. 

The transfer of energy and momentum in the $z$-direction between the field and the object are computed in the same manner as in the stationary case \cite{Vavilin2023, fernandez2017}, but with fields boosted in the direction opposite to the movement of the object $L_z(-\xi) \ket f$. It is important to note, that the wave function $\ket f$ is defined analytically in the plane wave basis and that it has significant angular momentum components beyond the $j_\text{max}$ found for the object at rest (Fig.~\ref{fig:crossec_wide_klist}). Such $j>j_\text{max}$ components in the laboratory frame can produce $j\leq j_\text{max}$ components in the frame of the object [\Eq{eq:boostwfangular}]. It is therefore required that the Lorentz boost should be applied {\em before} the truncation of the maximal multipolar order in the wave function of the incident field. In practice, one boosts fields in the plane wave basis via \Eq{eq:boostpwwf}, which loses no information about the wave function. Only after this transformation the relevant (up to $j_\text{max}$) angular momentum components $f_{jm\lambda}(k)=\braket{kjm\lambda|L_\xi(z) | f}$ should be extracted and used for the interaction with the scatterer. Schematically, the recipe can be formulated as
\eq{
f_\lambda(\bm k) \xrightarrow[\text{to co-moving frame}]{\text{Boost}} f'_\lambda(\bm k) \xrightarrow[\text{change}]{\text{Basis}} f'_{jm\lambda}(k) \xrightarrow[\text{truncated}]{\text{T-matrix}} g'_{jm\lambda}(k),\label{eq:scheme}
}
where the last step may be equivalently substituted with the action of the $S$-matrix truncated at the same maximal multipolar order as the $T$-matrix.

\begin{figure}[h]
 \begin{subfigure}[b]{0.5\textwidth}
        \centering
	%\textbf{}
        \includegraphics[width=\textwidth]{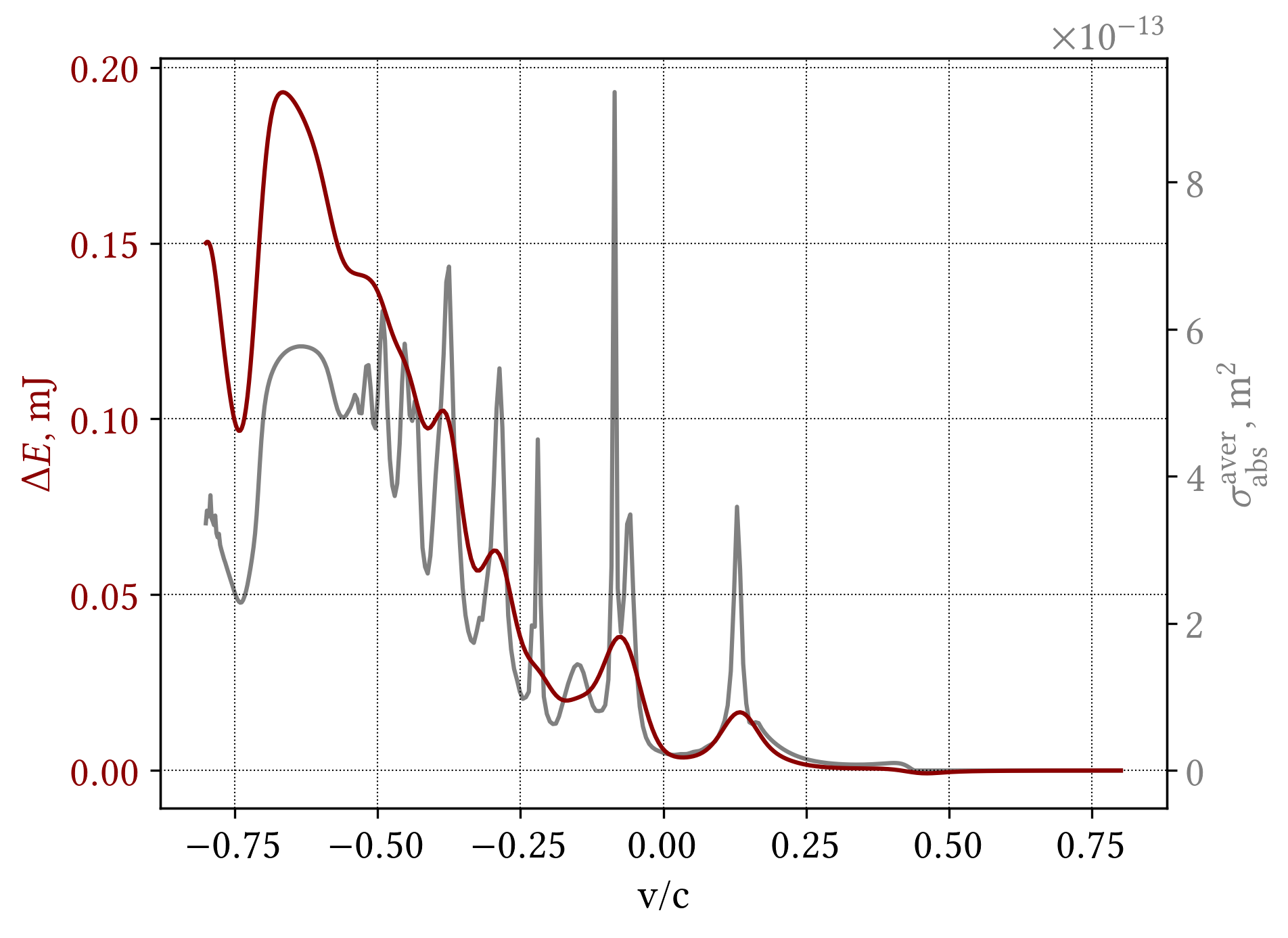}
        \caption{}
	\label{fig:nrj_transf_obj_frame}
     \end{subfigure}
     \hfill
     \begin{subfigure}[b]{0.5\textwidth}
         \centering
        \includegraphics[width=\textwidth]{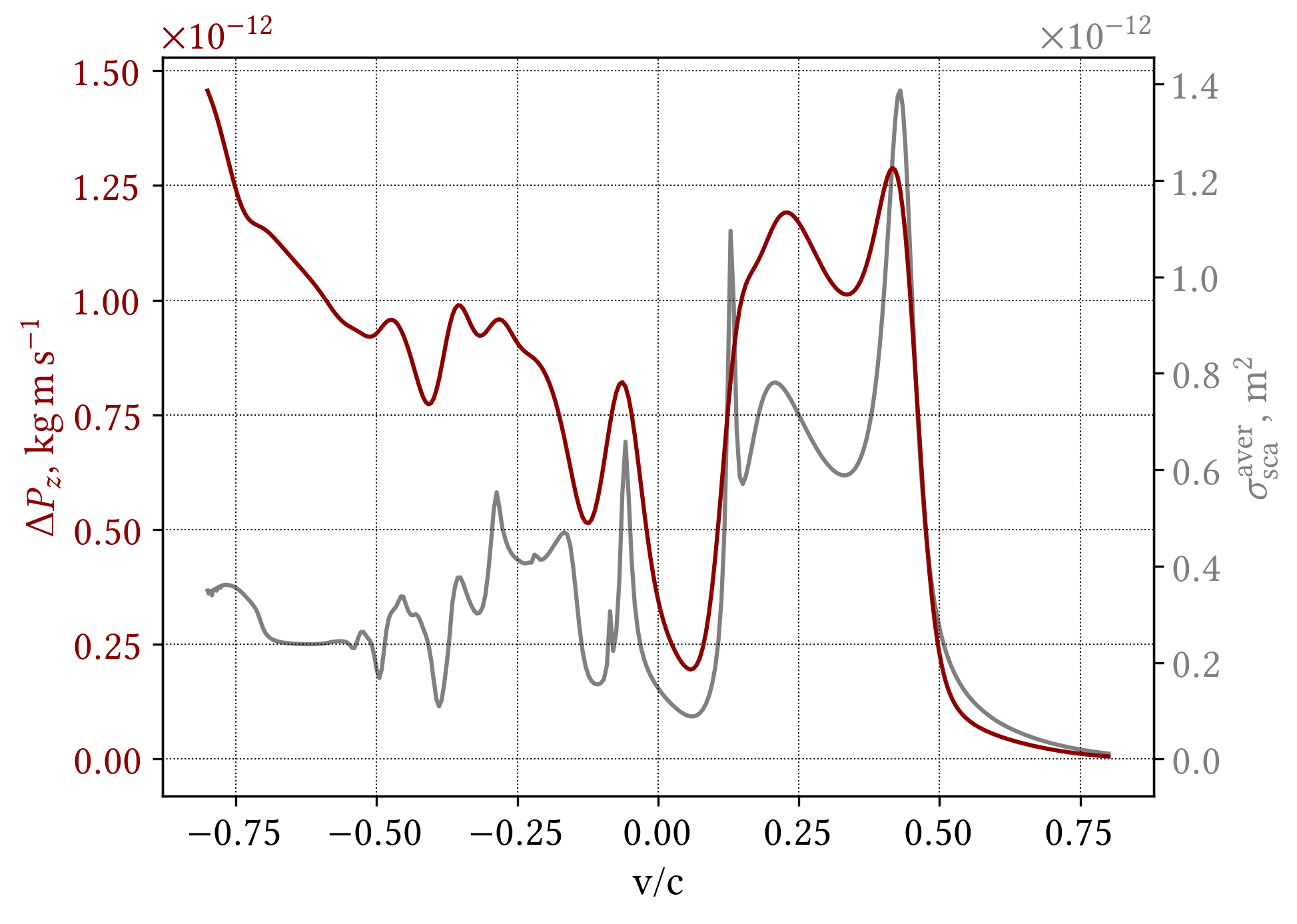}
        \caption{}
	\label{fig:momz_transf_obj_frame}
     \end{subfigure}
    % \end{centering}
    % }
	\caption{
		  The red lines show the transfer of (a) energy and (b) momentum in the $z$-direction from the electromagnetic pulse $\ket f$ to the silicon sphere that moves at different velocities along the z-axis, computed in the reference frame of the sphere. For the reference, the grey lines show (a) absorption and (b) total interaction cross-section of the silicon sphere is given, evaluated at wave number corresponding to the Doppler-shifted peak of the spectrum of the pulse. A positive sign of $v/c$ corresponds to the sphere moving in the positive $z$-direction of the laboratory frame.
		}\label{fig:transf_obj}
\end{figure}

The results for the transfer of energy and momentum using the expressions in (\ref{eq:transferhpz}) are depicted in Fig.~(\ref{fig:transf_obj}). For comparison, in grey, the rotationally averaged absorption and scattering cross-sections  
\eq{
\sigma^\text{aver}_\text{abs}(k_p) &= \frac{4\pi}{k^2}\text{Trace}\big[\mathds 1 - S^\dagger(k_p) S(k_p)\big]\\
\sigma^\text{aver}_\text{sca}(k_p) &= \frac{4\pi}{k^2}\text{Trace}\big[T^\dagger(k_p) T(k_p)].
} 
of the silicon sphere at rest are presented, evaluated at wave numbers corresponding to the Doppler shifted peak of the spectrum of the pulse. For each velocity, the center wavelength of the pulse $k_0$ shifts as
\eq{
k_p &\approx k_0 (\cosh(\xi) - \cos(0)\sinh(\xi)) \\
&= k_0 \,e^{-\xi} = k_0 \,e^{-\arctanh v/c},
}
where the approximation holds because the incident pulse is focused along the positive z-axis. The pulse interacts with the sphere in the wave number domain around the new center wavelength, which makes the interaction stronger or weaker depending on the velocity. The shape of the energy transfer reflects the absorption profile, as expected. The shape of the momentum transfer $\Delta P_z$ can be compared to the profile of the averaged scattering cross-section, since the latter, in contrast to the average absorption, contains the scattering contribution.

The number of equidistant discretization points for the wave function in the plane wave basis $f_\lambda(\bm k)$ is $N_k = 200$, $N_\theta = 200$, $N_\phi = 100$. Boosts in the plane wave basis Eq.~(\ref{eq:boostpwwf}) consist in changing the assignment of the initial values to the new domain points $(k'(\theta), \theta' , \phi')$ according to Eqs.~(\ref{eq:plboosttheta}-\ref{eq:plboostphi}). The domain stops being rectangular and equidistant in $k$ and $\theta$. After extracting the required angular momentum components of the boosted wave function, one must use the T-matrix that corresponds to the new wave number domain. This allows computation of outgoing coefficients via Eq.~(\ref{eq:inout2}). The final result is evaluated for rapidities between $\xi_\text{min}=-1$ and $\xi_\text{max}=1$ in equidistant steps for $N_\xi = 400$ points.

\subsection{Laboratory frame of reference}
\begin{figure}[h]
 \begin{subfigure}[b]{0.5\textwidth}
        \centering
	%\textbf{}
        \includegraphics[width=\textwidth]{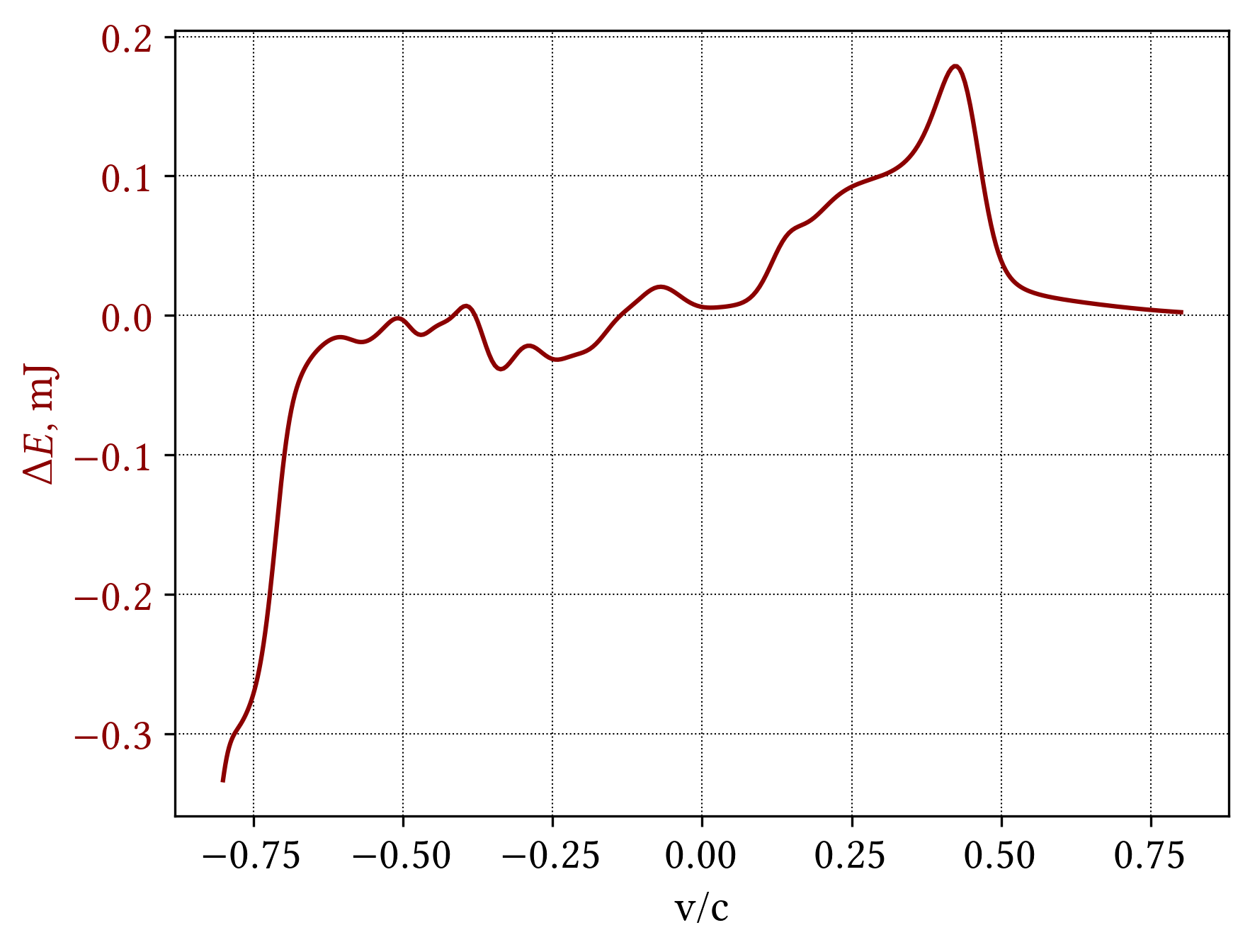}
        \caption{}
	\label{fig:nrj_transf_lab_frame}
     \end{subfigure}
     \hfill
     \begin{subfigure}[b]{0.5\textwidth}
         \centering
        \includegraphics[width=\textwidth]{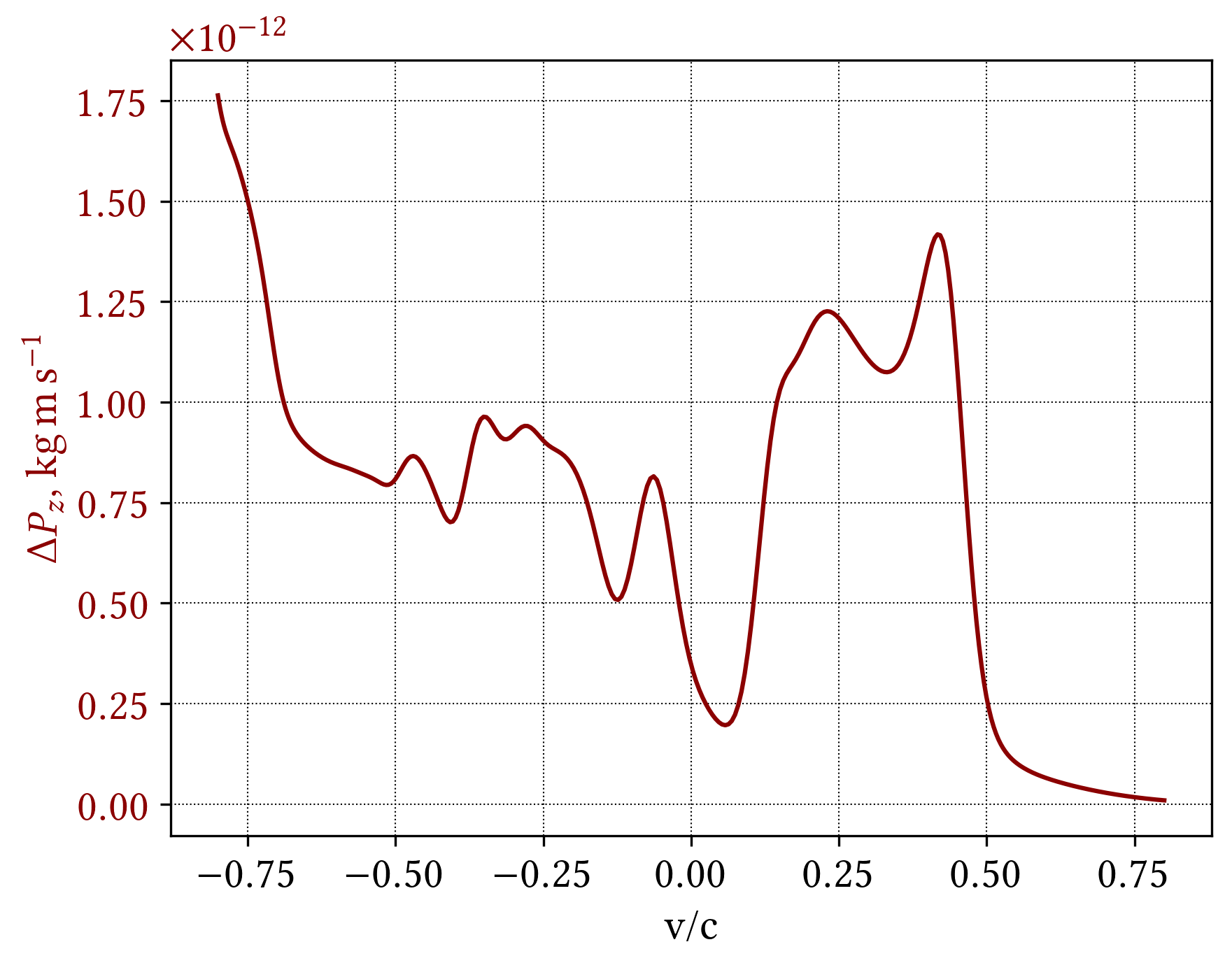}
        \caption{}
	\label{fig:momz_transf_lab_frame}
     \end{subfigure}
    % \end{centering}
    % }
	\caption{
		Loss of the (a) energy and (b) momentum in the $z$-direction from the electromagnetic pulse $\ket f$ when scattered by a silicon sphere that moves at different velocities along the z-axis, as observed in the laboratory frame. The positive sign of $v/c$ corresponds to the sphere moving in positive $z$-direction 
		}\label{fig:transf_lab}
\end{figure}

Now the transfer of quantities is computed in the laboratory frame of reference. We use the transformation properties of generators responsible for the corresponding quantities: $H = c P^0$ for energy and $P_z=P^3$ for momentum in the $z$-direction:
\eq{
	L_z(\xi) H L^{-1}_z(\xi) &= \cosh(\xi) H + \sinh(\xi) P_z c \\
	L_z(\xi) P_z L^{-1}_z(\xi) &= \sinh(\xi) H/c + \cosh(\xi) P_z,
}
which implies the connection between the scalar product values in two frames to be
\eq{
	\braket{\Delta H}^\text{lab} &= \cosh(\xi) \braket{\Delta H}^\text{obj} - \sinh(\xi) \braket{\Delta P_z}^\text{obj} c \\
	\braket{\Delta P_z}^\text{lab} &= -\sinh(\xi) \braket{\Delta H}^\text{obj}/c + \cosh(\xi) \braket{\Delta P_z}^\text{obj}.
	\label{eq:changeframes}
}
The transfer of energy and momentum in the laboratory frame is shown in Fig.~(\ref{fig:transf_lab}).

A notable phenomenon can be observed for some regions of negative $v$, when the object moves towards the pulse. In the laboratory frame of reference, the energy of the electromagnetic field increases. We note that, in contrast, this does not happen in the frame of reference of the object (see Fig.~\ref{fig:nrj_transf_obj_frame}). In a first order approximation, the phenomenon can be attributed to the Doppler effect. Assuming now a monochromatic illumination for simplicity, one readily appreciates that the Doppler effect can increase the energy as measured in the laboratory frame, because when the object moves towards the source, the frequency of the scattered field coming back to the source will be larger than the one emitted by the source. For the more complicated case of the polychromatic pulse, one can apply the previous argument to each of the frequencies composing the spectrum. A more detailed analysis, which we do not perform here, would take into account the absorption of energy by the material and the directionality of the scattering.

\subsection{Polychromatic T-matrix of the moving silicon sphere}
The scattering in the laboratory reference frame can also be described via the boosted T-matrix, which acts on the incident field $\ket f$ that is defined in the laboratory frame. In this section we numerically compute the Lorentz boosted T-matrix of the silicon sphere and use it to compute the transfer of momentum in the laboratory frame. 

The matrix element of the Lorentz boost in angular momentum basis is \cite[Sec. 2.2.1]{Vavilin2023}
\eq{
&\braket{k' j' m' \lambda' | L_z(\xi) | k j m \lambda}= \nonumber \\
&\qquad=\delta_{\lambda' \lambda} \delta_{m' m} \Theta\big( \abs{\xi} -  \abs{\ln(k'/k)}\big) \frac{\sqrt{2j'+1} \sqrt{2j+1}}{2 k' k \sinh (\abs{\xi})} d^{j'}_{m\lambda}(\theta')  d^j_{m\lambda}(\theta)\label{eq:lor-matelem},
}
which can be used to transform T-matrices via $\widetilde T = L_z(\xi) T L^{-1}_z(\xi)$. The result of the application is (see App.~\ref{sec:boost_diag_tmat}):
\eq{
&\widetilde T^{j_1 m_1 \lambda_1}_{j_2 m_2 \lambda_2}(k_1, k_2) =  \int_{k_1 e^{-\abs{\xi}}}^{k_1 e^{\abs{\xi}}} dk_2' \, k_2' \sum_{j'_1=1}^\infty \sum_{j'_2=1}^\infty \, T^{j'_1 m_1 \lambda_1}_{j'_2 m_2 \lambda_2}(k_2')\,\times\nonumber \\
&\times \frac{1}{2}\,\sqrt{2j_1+1}\,\sqrt{2j_1'+1}\, \frac{1}{k_1 k_1' \abs{\sinh\xi}}\, d^{j_1}_{m_1 \lambda_1}(\theta) \, d^{j_1'}_{m_1 \lambda_1}(\theta') \nonumber \\
&\times \frac{1}{2}\,\sqrt{2j_2'+1}\,\sqrt{2j_2+1}\, \frac{\Theta\big(\abs{\xi} - \abs{\ln(k_2'/k_2)} \big)}{k_2' k_2 \abs{\sinh\xi}}\,d^{j_2}_{m_2 \lambda_2}(\theta_2)\, d^{j_2'}_{m_2 \lambda_2}(\theta_2'),
}
with
\eq{
\cos \theta = \frac{k_1 \cosh \xi - k_2'}{k_1 \sinh \xi},\qquad \cos \theta' = \frac{k_1 - k_2' \cosh \xi}{k_2' \sinh \xi},\\
\cos \theta_2 = -\frac{k_2' - k_2 \cosh \xi}{k_2 \sinh \xi}, \qquad \cos \theta_2' = -\frac{k_2' \cosh \xi - k_2}{k_2' \sinh \xi}.
}

\begin{figure}
\centering
% First subfigure
\begin{subfigure}[b]{\textwidth}
    \centering
    \includegraphics[width=\textwidth]{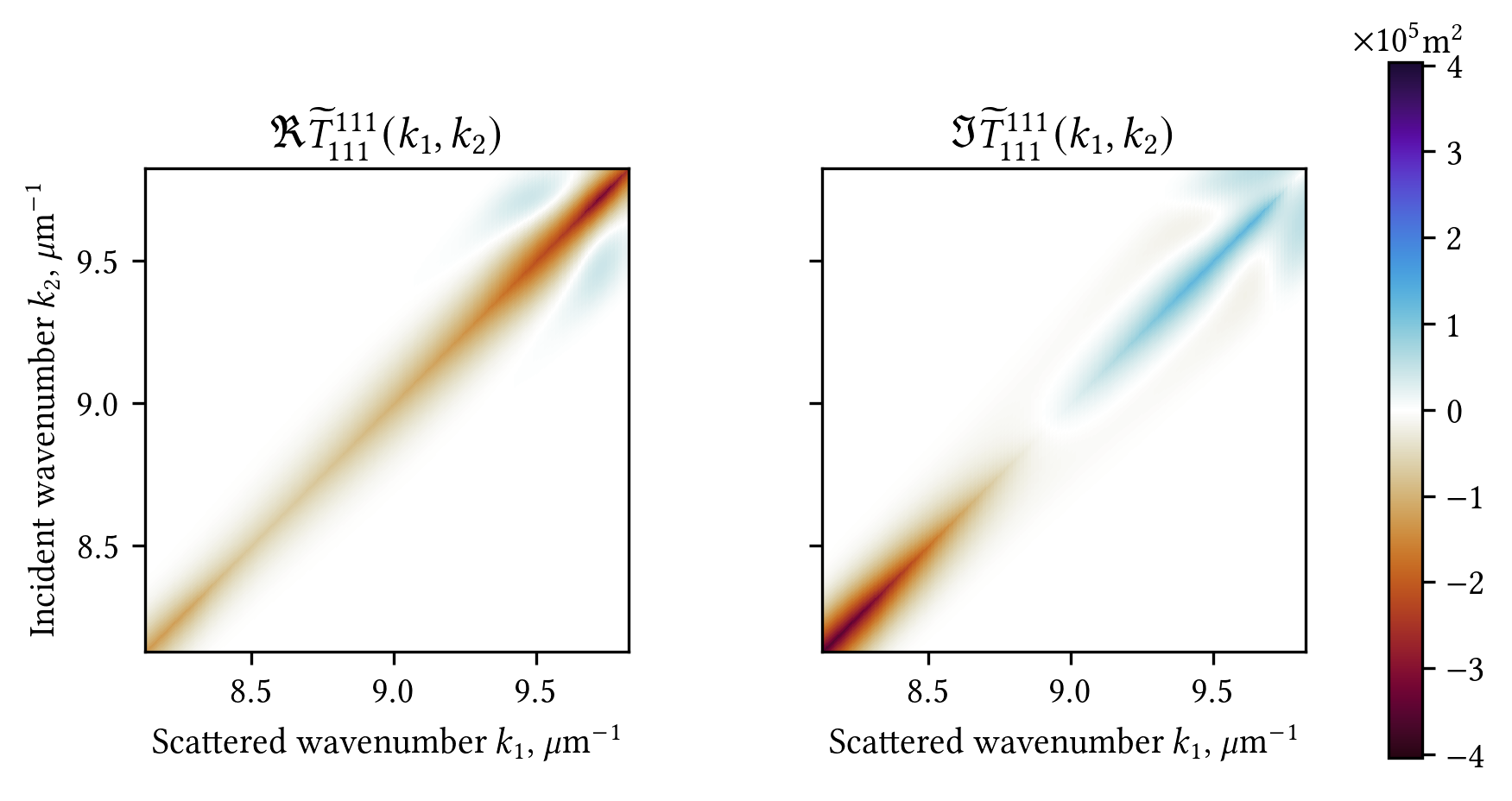}
    \caption{$\xi=0.025$}
    \label{fig:sub1}
\end{subfigure}
\par\bigskip % Adds space between the subfigures

% Third subfigure
\begin{subfigure}[b]{\textwidth}
    \centering
    \includegraphics[width=\textwidth]{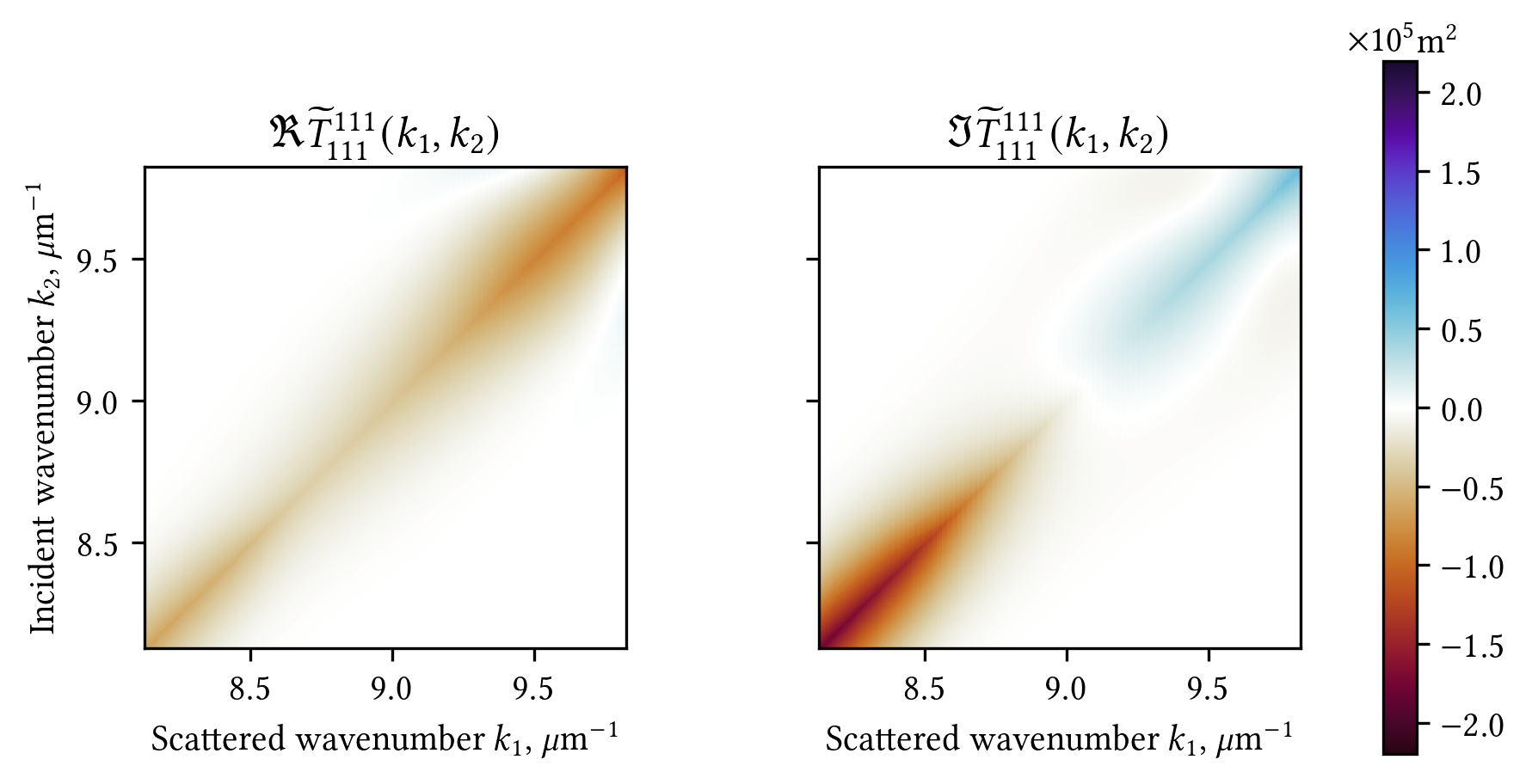}
    \caption{$\xi=0.05$}
    \label{fig:sub3}
\end{subfigure}

\caption{Elements of the polychromatic T-matrix for the silicon sphere moving with different boost parameters $\xi = \arctanh(v/c)$. The real part of the T-matrix element is shown on the left and the imaginary part is on the right. The higher the velocity, the wider the spreading of the scattered wave number for a fixed incident wave number.
}
\label{fig:BoostedTmat}
\end{figure}

One may see that Lorentz boost of the frequency-diagonal T-matrix of the silicon sphere makes it non-diagonal in frequency, as expected for moving objects. The element of the boosted T-matrix with $j_{1,2}=1$, $m_{1,2}=1$, $\lambda_{1,2}=1$, as a function of incident wave number $k_1$ and scattered wave number $k_2$ is illustrated in Fig.~(\ref{fig:BoostedTmat}),  computed for rapidity $\xi=0.025$ and 0.05.

We compute the difference of momentum between the outgoing and incoming fields similar to the stationary case
\eq{
\braket{\Delta P_z} &= \braket {f|P_z|f} - \braket {h|P_z|h},
}
but now the computation of the outgoing field involves integration over the incident wave number:
\eq{
	h_{jm\lambda}(k) = f_{jm\lambda}(k) + \int_0^\infty dk' \, k' \,  \sum_{\lambda' = \pm 1}\sum_{j'=1}^{j_\text{max}}\sum_{m'=-j}^j\widetilde T^{j m \lambda}_{j' m' \lambda'}(k, k') f_{j'm'\lambda'}(k'). \label{eq:Tfpoly}
}
The integral in $k'$ can be truncated to the region of a significant part of the Gaussian wave number profile of $\ket f$.

\begin{figure}[h]
        \centering
	%\textbf{}
        \includegraphics[width=0.75\textwidth]{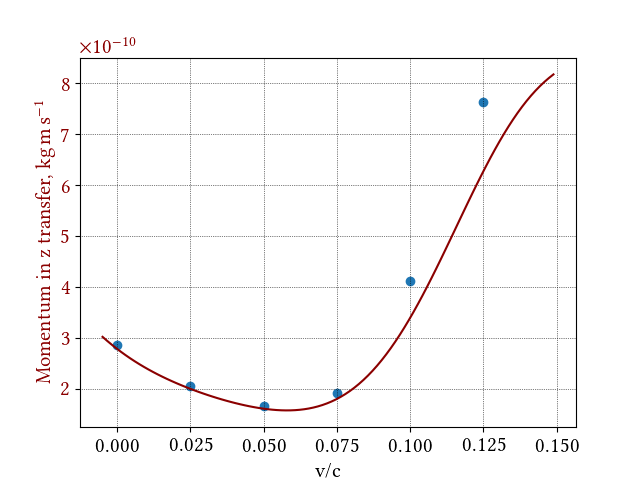}
	\caption{
		Transfer of momentum $P_z$ between the electromagnetic pulse $\ket f$ and the moving silicon sphere in the laboratory frame, computed via the polychromatic T-matrix (black) and via transformed quantities from the co-moving frame (blue) as reference.
		}\label{fig:PolyMomTransfer}
\end{figure}

Finally, the transfer of momentum is computed for a number of velocities $v/c=0.00001$, 0.025, 0.05, 0.075, 0.1 and 0.125. The results are depicted in Fig.~(\ref{fig:PolyMomTransfer}), next to the reference computed in the last section. The transfer via the boosted T-matrix in the lab frame fails for higher velocities, when the chosen $j_\text{max}=6$ cannot account completely for the scattering.

We note that the T-matrix of the moving object has significant elements for all values of total angular momentum $j$, because the interaction may happen arbitrarily far from the origin of the reference frame. Therefore, the range of the multipolar order can not be truncated without loss of the information about the scatterer. This contrasts sharply with the possibility of such lossless truncation for the T-matrix of a stationary object. Practically, this means that the value of $j_\text{max}$ in Eq.~(\ref{eq:Tfpoly}) is dictated by the character of the incident field. In particular, $j_\text{max}$ should encompass a region around the origin large enough to completely account for the spatial domain of the interaction between the field and the scatterer. For example, consider an interaction of a localized pulse hitting a moving object when they both are near the origin of the reference frame. A smaller $j_\text{max}$ is required in this case compared to the interaction of the same moving object with a pulse designed to hit the object further from the origin. Similarly, it is impossible to find a $j_\text{max}$ to completely describe the interaction of an ideal plane wave with a moving object, since there will be points in space, where the interaction is happening, and that are arbitrarily far from the origin. 
Computations in the co-moving frame may always be conducted with the $j_\text{max}$ dictated by the object's finite size and not the pulse. This makes it beneficial to compute quantities in the co-moving frame as depicted in \Eq{eq:scheme}, and to transform them to the laboratory frame if needed, as in \Eq{eq:changeframes}.

\section{Conclusions}\label{sec:concl}
We have applied the polychromatic T-matrix formalism to the interaction between light and relativistically moving objects. While the exemplary calculations are performed for a particular light pulse and a silicon sphere, the formulas and the procedure contained in this article apply to generic polychromatic illuminations and generic objects. The transfer of energy and momentum between the pulse and the silicon sphere, which is in constant uniform motion, has been computed for a large range of velocities via switching to the co-moving frame of reference, where the T-matrix of the object at rest is used. We have compared this method to the direct use of the Lorentz boosted T-matrix in the laboratory frame, where we computed the polychromatic T-matrix of a moving silicon sphere. The second method is found to be rather impractical because the interaction may happen arbitrarily far from the origin of the reference frame, and then, the multipolar orders of the light-matter interaction cannot really be truncated without losing information about the scatterer. One may, however, obtain any desired quantity in the laboratory frame by suitably transforming the one obtained in the object frame, as we have shown in the article.

The presented theoretical framework can be applied to light sails, including the practical optimization of sail designs and material selection to enhance the efficiency of the propulsion of the devices. Moreover, the application of the methodology to objects at rest has immediate applications to pulsed optical traps and tweezers.

The many publicly available resources for computing T-matrices \cite{TmatrixCodes} make our approach computationally friendly. The codes use the recently released \textsc{treams} package \cite{Beutel2023b}.

\section{Acknowledgments}
This work was partially funded by the Deutsche Forschungsgemeinschaft (DFG, German Research Foundation) -- Project-ID 258734477 -- SFB 1173. J.~D.~Mazo--V\'asquez is part of the Max Planck School of Photonics, supported by the German Federal Ministry of Education and Research (BMBF), the Max Planck Society, and the Fraunhofer Society.

\appendix
\section{Lorentz boost of frequency-diagonal T-matrix}
\label{sec:boost_diag_tmat}
The matrix element of the Lorentz boost in angular momentum basis is \cite[Sec. 2.2.1]{Vavilin2023}:
\eq{
&\braket{k' j' m' \lambda' | L_z(\xi) | k j m \lambda}= \nonumber \\
&\qquad=\delta_{\lambda' \lambda} \delta_{m' m} \Theta\big( \abs{\xi} -  \abs{\ln(k'/k)}\big) \frac{\sqrt{2j'+1} \sqrt{2j+1}}{2 k' k \sinh (\abs{\xi})} d^{j'}_{m\lambda}(\theta')  d^j_{m\lambda}(\theta)\label{eq:app-lor-matelem},
}
which can be used to transfrom T-matrices via $\widetilde T = L_z(\xi) T L^{-1}_z(\xi)$:
\eq{
&\widetilde T^{j_1 m_1 \lambda_1}_{j_2 m_2 \lambda_2}(k_1, k_2) = \braket{k_1 j_1 m_1 \lambda_1 | L_z(\xi)\, T\, L_z^{-1}(\xi) |k_2 j_2 m_2 \lambda_2} \\
&= \int_0^\infty dk_1' \, k_1' \int_0^\infty dk_2' \, k_2' \sum_{j'_1=1}^\infty \sum_{j'_2=1}^\infty \, T^{j'_1 m_1 \lambda_1}_{j'_2 m_2 \lambda_2}(k_1', k_2') \times\\
&\times \frac{1}{2}\,\sqrt{2j_1+1}\,\sqrt{2j_1'+1}\, \frac{\Theta\big(\abs{\xi} - \abs{\ln(k_1/k_1')} \big)}{k_1 k_1' \abs{\sinh\xi}}\, d^{j_1}_{m_1 \lambda_1}(\theta_1) \, d^{j_1'}_{m_1 \lambda_1}(\theta_1')  \\
&\times \frac{1}{2}\,\sqrt{2j_2'+1}\,\sqrt{2j_2+1}\, \frac{\Theta\big(\abs{\xi} - \abs{\ln(k_2'/k_2)} \big)}{k_2' k_2 \abs{\sinh\xi}}\, d^{j_2'}_{m_2 \lambda_2}(\theta_2') \, d^{j_2}_{m_2 \lambda_2}(\theta_2),
}
where $\theta_{1,2}$, $\theta_{1,2}'$ are defined via
\eq{
\cos \theta_1 = \frac{k_1 \cosh \xi - k_1'}{k_1 \sinh \xi},\qquad \cos \theta_1' = \frac{k_1 - k_1' \cosh \xi}{k_1' \sinh \xi},\\
\cos \theta_2' = -\frac{k_2' \cosh \xi - k_2}{k_2' \sinh \xi},\qquad \cos \theta_2 = -\frac{k_2' - k_2 \cosh \xi}{k_2 \sinh \xi}.
}
Since the T-matrix of the sphere is diagonal in frequency, this  general expression can be simplified. Using
\eq{
	 T^{j_1 m_1 \lambda_1}_{j_2 m_2 \lambda_2}(k_1, k_2) = T^{j_1 m_1 \lambda_1}_{j_2 m_2 \lambda_2}(k_2) \,\frac{1}{k_2}\, \delta(k_1 - k_2),
}
the transformed T-matrix is written as
\eq{
&\widetilde T^{j_1 m_1 \lambda_1}_{j_2 m_2 \lambda_2}(k_1, k_2) = \\
&= \int_0^\infty dk_1' \, k_1' \int_0^\infty dk_2' \, k_2' \sum_{j'_1=1}^\infty \sum_{j'_2=1}^\infty \, T^{j'_1 m_1 \lambda_1}_{j'_2 m_2 \lambda_2}(k_2')\,\frac{1}{k_2'}\, \delta(k_1' - k_2') \times\nonumber \\
&\times \frac{1}{2}\,\sqrt{2j_1+1}\,\sqrt{2j_1'+1}\, \frac{\Theta\big(\abs{\xi} - \abs{\ln(k_1/k_1')} \big)}{k_1 k_1' \abs{\sinh\xi}}\, d^{j_1}_{m_1 \lambda_1}(\theta_1) \, d^{j_1'}_{m_1 \lambda_1}(\theta_1') \nonumber \\
&\times \frac{1}{2}\,\sqrt{2j_2'+1}\,\sqrt{2j_2+1}\, \frac{\Theta\big(\abs{\xi} - \abs{\ln(k_2'/k_2)} \big)}{k_2' k_2 \abs{\sinh\xi}}\, d^{j_2'}_{m_2 \lambda_2}(\theta_2') \, d^{j_2}_{m_2 \lambda_2}(\theta_2).
}
Integration over the wave number eliminates $k_1$ and one of the Heaviside functions truncates the integral in $k_2$:
\eq{
&\widetilde T^{j_1 m_1 \lambda_1}_{j_2 m_2 \lambda_2}(k_1, k_2) =  \int_0^\infty dk_2' \, k_2' \sum_{j'_1=1}^\infty \sum_{j'_2=1}^\infty \, T^{j'_1 m_1 \lambda_1}_{j'_2 m_2 \lambda_2}(k_2')\,\times \nonumber\\
&\times \frac{1}{2}\,\sqrt{2j_1+1}\,\sqrt{2j_1'+1}\, \frac{\Theta\big(\abs{\xi} - \abs{\ln(k_1/k_2')} \big)}{k_1 k_1' \abs{\sinh\xi}}\, d^{j_1}_{m_1 \lambda_1}(\theta) \, d^{j_1'}_{m_1 \lambda_1}(\theta')\nonumber  \\
&\times \frac{1}{2}\,\sqrt{2j_2'+1}\,\sqrt{2j_2+1}\, \frac{\Theta\big(\abs{\xi} - \abs{\ln(k_2'/k_2)} \big)}{k_2' k_2 \abs{\sinh\xi}}\, d^{j_2'}_{m_2 \lambda_2}(\theta_2') \, d^{j_2}_{m_2 \lambda_2}(\theta_2) \\
&=  \int_{k_1 e^{-\abs{\xi}}}^{k_1 e^{\abs{\xi}}} dk_2' \, k_2' \sum_{j'_1=1}^\infty \sum_{j'_2=1}^\infty \, T^{j'_1 m_1 \lambda_1}_{j'_2 m_2 \lambda_2}(k_2')\,\times\nonumber \\
&\times \frac{1}{2}\,\sqrt{2j_1+1}\,\sqrt{2j_1'+1}\, \frac{1}{k_1 k_1' \abs{\sinh\xi}}\, d^{j_1}_{m_1 \lambda_1}(\theta) \, d^{j_1'}_{m_1 \lambda_1}(\theta') \nonumber \\
&\times \frac{1}{2}\,\sqrt{2j_2'+1}\,\sqrt{2j_2+1}\, \frac{\Theta\big(\abs{\xi} - \abs{\ln(k_2'/k_2)} \big)}{k_2' k_2 \abs{\sinh\xi}}\,d^{j_2}_{m_2 \lambda_2}(\theta_2)\, d^{j_2'}_{m_2 \lambda_2}(\theta_2'),
}
with
\eq{
\cos \theta = \frac{k_1 \cosh \xi - k_2'}{k_1 \sinh \xi},\qquad \cos \theta' = \frac{k_1 - k_2' \cosh \xi}{k_2' \sinh \xi},\\
\cos \theta_2 = -\frac{k_2' - k_2 \cosh \xi}{k_2 \sinh \xi}, \qquad \cos \theta_2' = -\frac{k_2' \cosh \xi - k_2}{k_2' \sinh \xi}
}

\printbibliography[heading=bibintoc]

\end{document}